# Stability of Rotating Viscous and Inviscid flows

Hua-Shu Dou

Temasek Laboratories,
National Wind Tunnel Building
National University of Singapore,
Singapore 117508, SINGAPORE
Email: tsldh@nus.edu.sg; huashudou@yahoo.com

**Abstract:** Flow instability and turbulent transition can be well explained using a new proposed theory--Energy gradient theory [1]. In this theory, the stability of a flow depends on the relative magnitude of energy gradient in streamwise direction and that in transverse direction, if there is no work input. In this note, it is shown based on the energy gradient theory that inviscid non-uniform flow is unstable if the energy in transverse direction is not constant. This new finding breaks the classical linear theory from Rayleigh that inviscid flow is unstable if the velocity profile has an inflection point in parallel flows and inviscid flow is stable if the velocity profile has no inflection point in parallel flow. Then, stability of rotating viscous and inviscid flows is studied, and two examples of rotating flows (rotating rigid body motion and free vortex motion) are shown, respectively.

## 1. Introduction

In the classical theory for flow instability, Rayleigh (1880)[2] first developed a general linear stability theory for inviscid plane-parallel shear flows, and showed that a necessary condition for instability is that the velocity profile has a point of inflection. Later, Tollmien (1935) succeeded in showing that this also constitutes a sufficient condition for the amplification of disturbances [3]. Fjϕrtoft (1950) gave a further necessary condition for inviscid instability, that there is a maximum of vorticity for instability [4]. Therefore, it is well known that inviscid flow with inflectional velocity profile is unstable, while inviscid flow with no inflectional velocity profile is stable [5-8]. The associated analysis showed that the effect of viscosity is complex, and it plays dual roles to the flow instability. Recently, Dou (2002) [1] proposed a new theory for flow instability and turbulent transition—energy gradient theory. This new theory obtains good consistent results with the experiments for parallel flows and Taylor-Couette flows between two rotating cylinders. Using this theory, it is demonstrated that parallel shear flows with inflectional velocity profile for two-dimensional and axisymmetrical viscous flows are sufficient for instability [9]. On further studying the classical linear theory, it is found that there are some contradictions to the experimental results in this theory and these could not be reasonably explained so far. In this paper, the energy gradient theory is used to study the stability of rotating viscous and inviscid flows. It is demonstrated that inviscid non-uniform flow is unstable if the energy in transverse direction is not constant. This result overthrows the classical Rayleigh theorem on instability of inflectional velocity profile. Then, stability of rotating viscous and inviscid flows is studied, and two examples of rotating flows (rotating rigid body motion and free vortex motion) are shown, respectively.

## 2. Energy gradient theory

Recently, Dou [1] proposed an *energy gradient theory* with the aim to clarify the mechanism of transition from laminar to turbulence for wall bounded shear flows. Here, we give a short discussion for a better understanding of the work presented in this study. In the theory, the whole flow field is treated as an energy field. It is thought that the gradient of total energy in the transverse direction of the main flow and the viscous friction in the streamwise direction dominate the instability phenomena and hence the flow transition for a given disturbance. The energy gradient in the transverse direction has the potential to amplify a velocity disturbance, while the viscous friction loss in the streamwise direction can resist and absorb this disturbance. The flow instability or the transition to turbulence depends on the relative magnitude of these two roles of energy gradient amplification and viscous friction damping to the initial disturbance. Based on such, a new dimensionless parameter, $K$ (the ratio of the energy gradient in the transverse direction to that in the streamwise direction), is defined to characterize the stability of the base flow,

$$K = \frac{\partial E / \partial n}{\partial E / \partial s}. \qquad (1)$$

Here, $E = p + \frac{1}{2}\rho V^2 + \rho g \xi$ is the total energy for incompressible flows with $\xi$ as the coordinate in the direction of gravitational field, $n$ denotes the direction normal to the streamwise direction and $s$ denotes the streamwise direction. Furthermore, $\rho$ is the fluid density, g is the gravity acceleration, V is the velocity, and p is the hydrodynamic pressure. For pressure driven flows, the magnitude of energy gradient $\partial E / \partial s$ equals to the energy loss of unit volume fluid $\partial H / \partial s$ in unit length along the streamline due to the viscous friction. In other word, the mechanism of generation of streamwise energy gradient is resulted by the energy loss due to viscous friction (Fig.1). Therefore, for shear driven flows, the calculation of K can be obtained by the ratio of the energy gradient in transverse direction and the energy loss in unit length along the streamline [10],

$$K = \frac{\partial E / \partial n}{\partial H / \partial s}. \qquad (2)$$

As such, the parameter $K$ in Eq.(1) is a field variable. Thus, the distribution of $K$ in the flow field and the property of disturbance may be the perfect means to describe the disturbance amplification or decay in the flow. It is suggested that the flow instability can first occur at the position of $K_{max}$ which is construed to be the most "dangerous" position. Thus, for a given disturbance, the occurrence of instability depends on the magnitude of this dimensionless parameter *K* and the critical condition is determined by the maximum value of *K* in the flow. For a given flow geometry and fluid properties, when the maximum of *K* in the flow field is larger than a critical value $K_c$, it is expected that instability can occur for certain initial disturbance [1]. Turbulence transition is a local phenomenon in the earlier stage. For a given flow, K is proportional to the global Reynolds number. A large value of K has the ability to amplify the disturbance, and vice versa. The analysis has suggested that the transition to turbulence is due to the energy gradient and the disturbance amplification, rather than just the linear eigenvalue instability type as stated in [11,12]; this also follows that the pipe Poiseuille flow is linearly stable for all the Reynolds number. Both Grossmann [11] and Trefethen et al. [12] commented that the nature of the onset-of-turbulence mechanism in parallel shear flows must be different from an eigenvalue instability of linear equations of small disturbance. In fact, finite disturbance is needed for the turbulence initiation in the range of finite Re as found in experiments [13]. Dou [1] demonstrated that the criterion obtained has a consistent value at the subcritical condition of transition determined by the experimental data for plane Poiseuille flow, pipe Poiseuille flow as well as plane Couette flow. From this table it can be deduced that the turbulence transition takes place at a consistent critical value of $K_c$ at about 385 for both the plane Poiseuille flow and pipe Poiseuille flow, and about 370 for plane Couette flow. This may suggest that the subcritical transition in parallel flows takes place at a value of $K_c \approx 370\text{-}385$. The finding further suggests that the flow instability is likely resulted from the action of energy gradients, and not due strictly to the eigenvalue instability of linear equations. The critical condition for flow instability as determined by linear stability analysis differs largely from the experimental data for all the three different types of flows. Using energy gradient theory, it is also demonstrated that the viscous flow with an inflectional velocity profile is unstable for both two-dimensional flow and axisymmetrical (Fig.2)[9].

For plane Poiseuille flow, this said position where $K_{max} > K_c$ should then be the most dangerous location for flow breakdown, which has been confirmed by Nishioka et al's experiment [14]. The measured instantaneous velocity distributions indicate that the first oscillation of the velocity occurs at y/h=0.50~0.62.

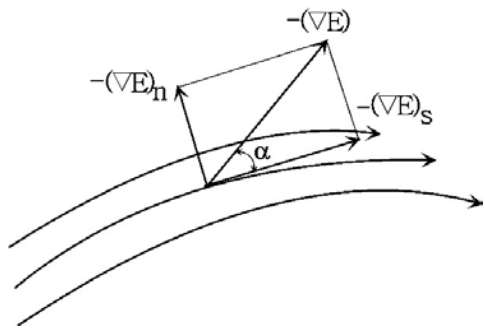 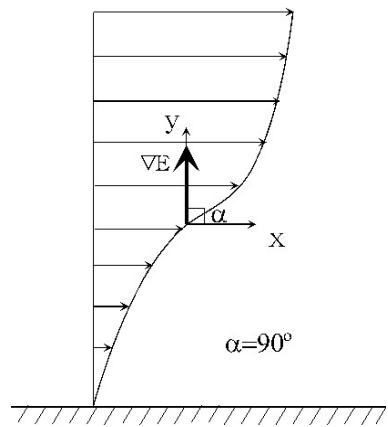

Fig. 1  Flow of any velocity profile.                Fig. 2 Velocity profile with an inflection point.

For pipe flow, in a recent study, Wedin and Kerswell [15] showed that there is the presence of the "shoulder" in the velocity profile at about r/R=0.6 from their solution of the traveling waves. The location of the "shoulder" is about same as that for $K_{max}$. According to the present theory, this "shoulder" may then be intricately related to the distribution of energy gradient. The solution of traveling waves has been confirmed by experiments recently [16].

As mentioned above, the mechanism for instability described by the parameter K is that it represents the balance of two roles of disturbance amplification by the energy gradient in transverse direction and disturbance damping by the energy loss in streamwise direction.

## 3. Stability of Viscous Rotating Flows

For rotating flows with axisymmetry, the energy gradient in circumferential direction is zero. The fluid flow is driven by external work input, otherwise the fluid is static. In this case the denominate in Eq.(1) is calculated by Eq.(2).

The following equation for calculating the radial distribution of energy loss along the streamline for rotating axisymmetric flow has been obtained as [17],

$$\frac{dH}{ds} \equiv \frac{\tau}{u}\frac{du}{dr} + \frac{\tau}{r} + \frac{d\tau}{dr}, \tag{3}$$

where the shear stress is

$$\tau = \mu\left(\frac{\partial u}{\partial r} - \frac{u}{r}\right). \tag{4}$$

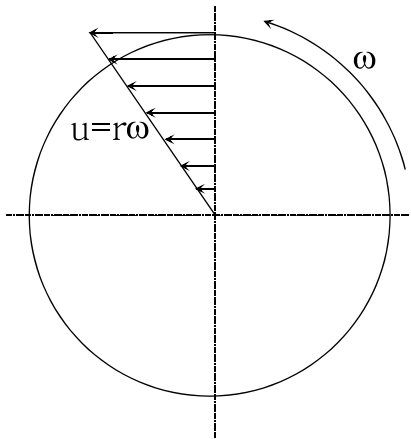
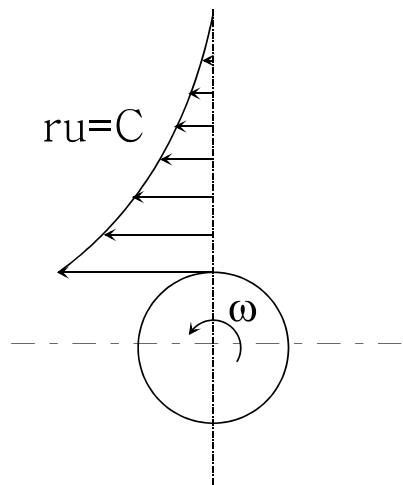

Fig. 3  Rigid body flow.　　　　　　　　Fig. 4  Free vortex flow.

The energy gradient in the transverse direction is

$$\frac{\partial E}{\partial r} = \frac{\partial (p + 1/2\rho u^2)}{\partial r} = \rho u \frac{du}{dr} + \rho \frac{u^2}{r}. \tag{5}$$

The parameter K is as

$$K = \frac{\partial E / \partial n}{\partial H / \partial s} = \frac{\rho u \dfrac{du}{dr} + \rho \dfrac{u^2}{r}}{\dfrac{\tau}{u}\dfrac{du}{dr} + \dfrac{\tau}{r} + \dfrac{d\tau}{dr}}. \tag{6}$$

The stability of the flow depends on the magnitude of this parameter K.

### 4. Stability of Inviscid Rotating flows

For inviscid flows, if there is no work input, the total energy along the streamline is constant. Thus, the total energy gradient in the streamwise direction for inviscid fluid is zero, $\partial E / \partial s = 0$. We can find that K is infinite from Eq.(1) if the total energy is not constant in the transverse direction. According to the energy gradient theory, the flow is unstable at this condition. For shear driven flows with work input (including rotating flows), the energy loss in streamwise direction for inviscid flows is zero. If the energy gradient in transverse direction is not zero, the flow will be unstable; if the energy gradient in transverse direction is zero, the flow stability in indefinite. At this case the flow stability depends on the disturbance.

### 5. Rotating Rigid Body Motion

The velocity distribution for the rigid body motion of fluid in a container (Fig.3) is

$$u = r\omega \text{ and } v = 0, \tag{7}$$

and the velocity gradient is

$$\frac{\partial u}{\partial r} = \omega. \tag{8}$$

The shear stress of viscous fluid of rotating flow is

$$\tau = \mu\left(\frac{\partial u}{\partial r} - \frac{u}{r}\right) = 0, \tag{9}$$

and

$$\frac{\tau}{r} = 0, \ \frac{d\tau}{dr} = 0, \ \frac{\tau}{u}\frac{du}{dr} = 0. \tag{10}$$

Thus, the energy loss along the streamwise direction is

$$\frac{dH}{ds} \equiv \frac{\tau}{u}\frac{du}{dr} + \frac{\tau}{r} + \frac{d\tau}{dr} = 0. \tag{11}$$

The energy gradient in the transverse direction is

$$\frac{\partial E}{\partial r} = \frac{\partial(p + 1/2\rho u^2)}{\partial r} = \rho u\frac{du}{dr} + \rho\frac{u^2}{r} = 2\rho\omega^2 r. \tag{12}$$

Introducing Eq.(11) and Eq.(12) into Eq.(2), the parameter K is as follow,

$$K = \frac{\partial E/\partial n}{\partial H/\partial s} = \infty. \tag{13}$$

Therefore, the viscous flow of rigid body motion is unstable.

For inviscid flow, the energy loss of the rigid body motion in a container is

$$\frac{dH}{ds} \equiv \frac{\tau}{u}\frac{du}{dr} + \frac{\tau}{r} + \frac{d\tau}{dr} \equiv 0. \tag{14}$$

The energy gradient of inviscid fluid flow in the transverse direction is the same as Eq.(12).

Introducing Eq.(14) and Eq.(12) into Eq.(2), the energy gradient parameter K is

$$K = \frac{\partial E/\partial n}{\partial H/\partial s} = \infty. \tag{15}$$

Therefore, the inviscid flow of rigid body motion is unstable, which is similar to the viscous case.

## 6. Free-Vortex Flow

The velocity distribution for the free-vortex flow of fluid (Fig.4) is

$$ru = c \text{ or } u = \frac{c}{r} \text{ and } v = 0, \tag{16}$$

and the velocity gradient is

$$\frac{\partial u}{\partial r} = -\frac{c}{r^2}. \tag{17}$$

The shear stress of viscous fluid is

$$\tau = \mu\left(\frac{\partial u}{\partial r} - \frac{u}{r}\right) = \mu\left(-\frac{c}{r^2} - \frac{c}{r^2}\right) = -\mu\frac{2c}{r^2}. \tag{18}$$

Thus,
$$\frac{\tau}{r} = -\mu\frac{2c}{r^3}, \quad \frac{d\tau}{dr} = \mu\frac{4c}{r^3}, \quad \frac{\tau}{u}\frac{du}{dr} = \mu\frac{2c}{r^3}. \tag{19}$$

We have,
$$\frac{dH}{ds} \equiv \frac{\tau}{u}\frac{du}{dr} + \frac{\tau}{r} + \frac{d\tau}{dr} = \mu\frac{2c}{r^3} - \mu\frac{2c}{r^3} + \mu\frac{4c}{r^3} = \mu\frac{4c}{r^3}. \tag{20}$$

The energy gradient in the transverse direction is
$$\frac{\partial E}{\partial r} = \frac{\partial(p + 1/2\rho u^2)}{\partial r} = \rho u\frac{du}{dr} + \rho\frac{u^2}{r} = -\frac{c^2}{r^3} + \frac{c^2}{r^3} = 0. \tag{21}$$

Introducing Eq.(20) and Eq.(21) into Eq.(2), the energy gradient parameter K is,
$$K = \frac{\partial E/\partial n}{\partial H/\partial s} = 0. \tag{22}$$

Therefore, the viscous flow of free vortex motion is always stable.

The energy loss for the free-vortex flow of inviscid fluid in a container is
$$\frac{dH}{ds} \equiv \frac{\tau}{u}\frac{du}{dr} + \frac{\tau}{r} + \frac{d\tau}{dr} = 0. \tag{23}$$

The energy gradient in the transverse direction is the same as Eq.(21). Introducing Eq.(21) and Eq.(23) into Eq.(2), the energy gradient K is
$$K = \frac{\partial E/\partial n}{\partial H/\partial s} = \frac{0}{0}. \tag{24}$$

Therefore, the stability of free-vortex flow of inviscid fluid is indefinite. At this case, the flow stability depends on the disturbance. If there is no disturbance, the flow is stable. If the disturbance is sufficient large, the flow will be unstable.

## 7. Conclusions

In energy gradient theory, sufficient large value of transversal energy gradient has the tendency to amplify a disturbance and may lead to flow instability. Streamwise energy loss tries to damp the disturbance and delay the instability. A dimensionless field variable K is used to characterize the relative magnitude of disturbance amplification and damping as well as the extent of instability. When the maximum of K in the flow field is larger than a critical value Kc, it is expected that the instability occurs.

In this paper, the energy gradient theory is used to study the stability of viscous and inviscid rotating flows. It is demonstrated that inviscid non-uniform flow is unstable if the energy gradient in transverse direction is not zero. This result overthrows the classical Rayleigh theorem on instability of inflectional velocity profile in parallel flow. Then, stability of rotating viscous and inviscid flows is studied, and stability of two typical examples of rotating flows (rotating rigid body motion and free vortex motion) are shown, respectively. It is demonstrated that both of the viscous flow and inviscid flow of rigid body motion are unstable. The viscous flow of free vortex motion is always stable. The stability of free-vortex flow of inviscid fluid is indefinite. At this case, the flow stability depends on the disturbance. If there is no disturbance, the flow is stable. If the disturbance is sufficient large, the flow will be unstable.